\documentclass[preprint, aps, pre, eqsecnum,amsmath,amssymb,showpacs]{revtex4}              

\usepackage[final]{graphicx} 

\newcommand{\bean}{\begin{eqnarray}}
\newcommand{\eean}{\end{eqnarray}}
\newcommand{\bea}{\begin{eqnarray*}}
\newcommand{\eea}{\end{eqnarray*}}
\newcommand{\beq}{\begin{equation}}
\newcommand{\eeq}{\end{equation}}

\def\vereq#1#2{\lower3pt\vbox{\baselineskip1.5pt \lineskip1.5pt
\ialign{$\hfill##\hfil$\crcr#2\crcr\sim\crcr}}}

\begin{document}

\title{Standing wave oscillations in binary mixture convection: from onset via
symmetry breaking to period doubling into chaos}

\author{P.~Matura, D.~Jung, and M.~L\"{u}cke}
\affiliation{Institut f\"{u}r Theoretische Physik, Universit\"{a}t
des Saarlandes, D-66041~Saarbr\"{u}cken, Germany\\}

\date{\today}

\begin{abstract}
Oscillatory solution branches of the hydrodynamic field equations describing
convection in the form of a standing wave (SW) in binary fluid mixtures
heated from below are determined completely for several negative Soret
coefficients $\psi$. Galerkin as well as finite-difference simulations were 
used. They were augmented by simple control methods to 
obtain also unstable SW states. For sufficiently negative $\psi$ unstable SWs
bifurcate subcritically out of the quiescent conductive state. They become 
stable via a saddle-node bifurcation when
lateral phase pinning is exerted. Eventually their invariance under time-shift
by half a period combined with reflexion at midheight of the fluid layer gets
broken. Thereafter they terminate by undergoing a period-doubling cascade into
chaos.
\end{abstract}


\maketitle
\begin{figure}
\centerline{\includegraphics[height=12cm,width=8cm]{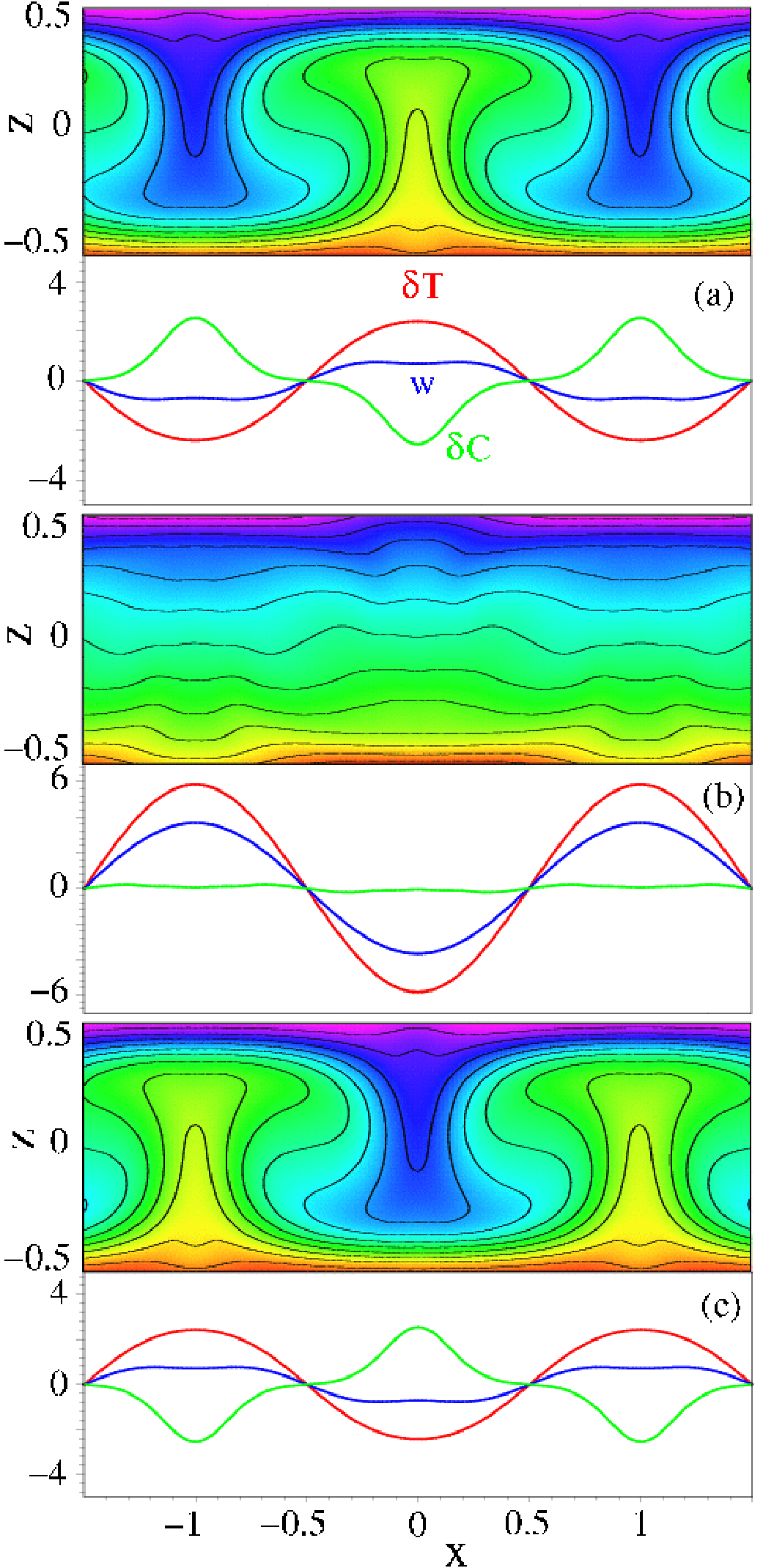}}
\caption{Snapshots of SW convection for $r=1.15,\,\psi= -0.25$ during
half of its oscillation period $\tau=2\pi/\omega=1.4$. The 
concentration distribution in the vertical cross-section through the layer 
is color coded with blue and red denoting high and low 
concentration, respectively. Lateral wave profiles of vertical velocity 
$w$, temperature $\delta T$, and concentration $\delta C$ at midheight, $z=0$, 
are shown by colored lines: $w$ - blue, 40\,$\delta T/R$ - red, and  
80\,$\delta C/R$ - green, respectively.
At the snapshot times $t=0$ (a), $0.265\,\tau$ (b), and $\tau/2$ (c)  
$\delta C(x=0,z=0,t)$ has a minimum, a zero crossing, and a 
maximum, respectively. This SW shows the MTS explained in the text.} 
\label{FIG:swpics}
\end{figure}

\clearpage 
Convection in one-component fluids like pure water occurs in Rayleigh-B\'enard 
setups of narrow channels heated from below in the form of stationary 
rolls. However, adding, say, 5\% of ethanol to the water the 
spatiotemporal behavior of the possible convective structures 
becomes much richer \cite{review}. The reason is that concentration variations 
which are generated via the Soret effect by 
externally imposed and by internal temperature gradients influence the buoyancy,
i.e., the driving force for convective flow. The latter in turn mixes by
advectively redistributing concentration. This nonlinear advection
gets in developed convective flow typically much larger than the smoothening 
by linear diffusion --- P\'eclet numbers measuring the strength
of advective concentration transport relative to diffusion are easily of the
order thousand. Thus, the concentration balance is strongly nonlinear 
giving rise to strong variations of the concentration field and to
boundary layer behavior as in Fig.~\ref{FIG:swpics}. In contrast to that, 
momentum and 
heat balances remain weakly nonlinear close to onset as in pure fluids implying
only smooth and basically harmonic variations of velocity and temperature fields
as of the critical modes, c.f. Fig.~\ref{FIG:swpics}.
Hence, the feedback interplay between ({\em i}) the Soret generated 
concentration variations, ({\em ii}) the resulting modified buoyancy, and
({\em iii}) the 
strongly nonlinear advective transport and mixing causes binary mixture 
convection to be rather complex not only with respect to its spatiotemporal
properties but 12 concerning its bifurcation behavior.

Take for example the case of negative Soret coupling, $\psi < 0$, 
between deviations $\delta T$ and $\delta C$ of temperature and concentration,
respectively, from their means \cite{coupling}. Then the above described cross 
coupling between solutal buoyancy and 
advection of Soret induced concentration variations generates oscillations.
They show up in transient growth of convection \cite{FL}
at supercritical heating, in relaxed nonlinear
traveling wave (TW) and standing wave (SW) solutions that branch subcritically
out of the conductive state via a common Hopf bifurcation, and in spatially
localized traveling wave (LTW) states. TW and LTW convection has been studied
experimentally and theoretically in detail 
\cite{review,Ahlers,Behringer,Kolodner,Platten,Steinberg,Surko,Yahata,Knobloch,JL02}. 
But little is known about 
nonlinear SW states beyond a weakly nonlinear amplitude equation analysis 
\cite{SZ93} that is restricted to the immediate vicinity of
the oscillatory threshold. It showed that SWs are unstable there, typically 
bifurcating backwards.

Here we determine for the first time structure, dynamics, and bifurcation 
behavior of SWs for several $\psi$. We stabilize the unstable SWs by 
control methods which can similarly also be applied in experiments. We found 
that they
undergo a period-doubling cascade into chaos after a mirror-timeshift symmetry
(c.f. further below) has been broken
that relates up- and downflow at a fixed lateral location, $x$, to each other.

Calculations were done for mixtures such as ethanol-water for Lewis number 
$L=0.01$ and Prandtl number $\sigma=10$. The field equations for convection 
were solved in a 
vertical cross section through the convection rolls perpendicular to their 
axes. A multi-mode Galerkin method as well as a finite-difference method was
used showing agreement with each other.
Horizontal boundaries at top and bottom, $z=\pm1/2$, were no-slip, perfectly 
heat conducting, and impermeable. Laterally we impose periodic boundary 
conditions with wave number $k=\pi$. In addition we suppress phase 
propagation. We stabilize the SW states by 
exerting control via the field amplitudes (or the heat current injected into 
the fluid) and the Rayleigh number $R$ in response to the 
instantaneous frequency and its temporal derivative, respectively
\cite{control}. In this way we
trace out the SW solution branch all the way from close to onset with large 
frequency to slowly oscillating SWs that eventually period-double into
chaos. The procedure starts from a supercritically 
growing transient SW with subsequent reduction of the heating below threshold.

We use $r=R/R_{c}^{0}$ to measure the thermal driving in terms of the
Rayleigh number reduced by the critical one $R_{c}^{0}=1707.762$ for 
onset of convection in a pure fluid. The flow induced mixing is measured
by the mixing number $M(t)$ that is defined in terms of the mean square 
concentration deviation, $M^2 = <(\delta C)^2>/<(\delta C_{cond})^2>$.  
 Here the brackets denote spatial averaging. In a perfectly mixed fluid 
 $M$ vanishes while $M=1$ in the 
 conductive state (denoted by the subscript $cond$) with its large 
 Soret-induced vertical concentration gradient.
 
In Fig.~\ref{FIG:swpics} we show snapshots of SW convection covering half an
oscillation period in order to display characteristic symmetry and 
structural properties. SWs are laterally mirror symmetric around positions
of maximal up- and downflow, e.g., $x$=0 and the node locations of the fields
are fixed in time. Furthermore, all fields have at every
instant definite parity under the mirror-glide operation $(x,z)\to
(x+\lambda/2,-z)$ of vertical reflection at mid-height, $z$=0, combined with 
lateral translation by half a wavelength. We did not observe SWs without
this symmetry -- perturbations breaking it that we introduced for test
purposes always decayed rapidly to zero. Finally, the fields of 
Fig.~\ref{FIG:swpics} have a 
definite  mirror-timeshift symmetry (MTS), e.g., $f(x,z,t)=-f(x,-z,t+\tau/2)$ 
for $f=\delta C,\delta T$, and the vertical velocity field $w$ with
$\tau=2\pi/\omega$ being the SW oscillation period. At mid-height the
condition $f(t)=-f(t+\tau/2)$ implies in particular that positive and negative
field extrema of an oscillation cycle have equal magnitudes.  
SWs with smaller frequency break this symmetry which is a prerequisite for 
period doubling \cite{SW84}.

Since the concentration balance is dominated by nonlinear advection the
distribution of $\delta C$ (color coded plots in Fig.~\ref{FIG:swpics}) shows
plume-like structures and narrow boundary layers. Consequently, the field
profiles of $\delta C$ which are shown in Fig.~\ref{FIG:swpics} at $z$=0 are
anharmonic. Also the temporal oscillation of $\delta C$ at a fixed location is
anharmonic. On the other hand, temporal and spatial variations of $w$ and 
$\delta T$ are much smoother and almost harmonic. The oscillation of $w$ is
temporally delayed relative to that of $\delta C$: the latter being
advected almost passively by the former changes the buoyancy driving force 
for $w$. At midheight this phase
shift increases from about 0.52$\pi$ at onset to about 0.73$\pi$ before the MTS
breaks.

\begin{figure}
\centerline{\includegraphics[height=12.5cm,width=8cm]{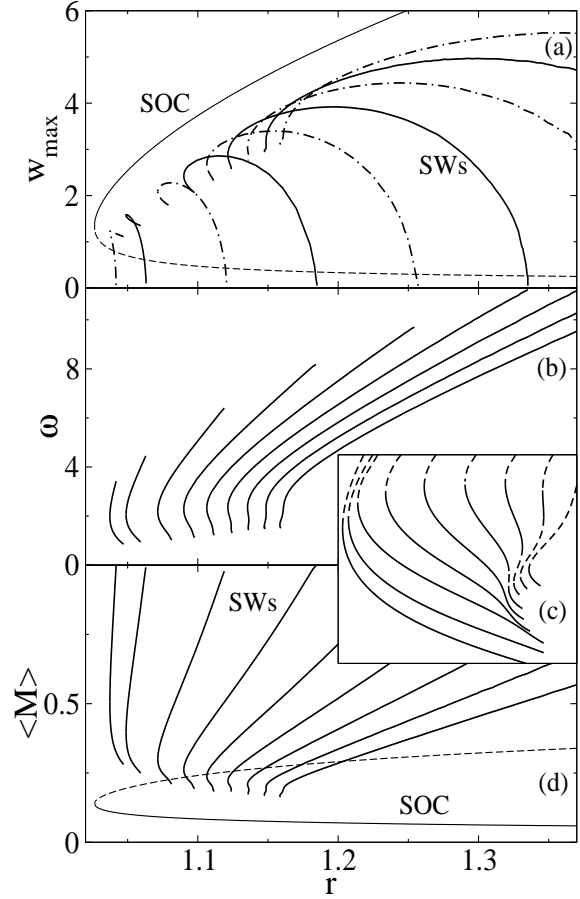}}
\caption{Bifurcation properties of SWs for  $\psi$= -0.03, -0.05, -0.1, -0.15, 
-0.2, -0.25, -0.3, -0.35, -0.4 (from left to right) : (a) Maximal 
vertical velocity $w_{max}$. (b) Frequency $\omega$. (d) Time average of 
the mixing number $M$. The inset (c) shows a blow up of the lower  
part of (b), however, with shifted solution branches to better display their
structural evolution with $\psi$. Full (dashed) lines in (c) identify stable
(unstable) SWs. Unstable SWs bifurcate subcritically out of the
quiescent conductive state [lower ends of the curves in (a); upper
ends in (b) and (c)] and undergo stability changes via saddle-node 
bifurcations . 
The SOC solution branch is shown for the sake of clarity only for $\psi$= -0.03.
SOC curves for the other $\psi$ are shifted slightly to the right.} 
\label{FIG:Bifall}
\end{figure}

Fig.~\ref{FIG:Bifall} shows how the bifurcation behavior of SWs changes
with Soret coupling strength $-0.4 \leq \psi \leq -0.03$. The solution branch 
for stationary overturning convection (SOC), which has the same spatial 
symmetries as SWs, is included for comparison only for $\psi=-0.03$.
The heating range in which SWs exist increases when $\psi$ becomes more 
negative since the oscillatory bifurcation threshold $r_{osc}$ is shifted 
stronger to higher $r$ than the SW saddle-node at $r_s^{SW}$ which marks the 
lower end of the  $r$-interval containing SWs. All these SWs bifurcate 
subcritically out of the conductive state as unstable solutions. They
become stable via saddle-node bifurcations. However, when the 
phase-pinning condition is lifted completely then SWs decay by developing TW 
transients since
any spatial phase difference between $\delta C$ and $w$ causes the extrema of
the latter to be "pulled" towards the solutally shifted buoyancy extrema.
Depending on $r$ these transients  either end in a nonlinear TW or SOC or the 
conductive state.

Moving along an SW branch the maximal 
vertical upflow velocity $w_{max}$ [Fig.~\ref{FIG:Bifall}(a)] does not
increase monotonically as in TWs and SOCs but rather has a maximum somewhat
below the respective SOC value before it drops again. On the other
hand, $\omega$ and $M$ decrease monotonically starting with the Hopf 
frequency $\omega_{H}$ and $M=1$, respectively, at onset at the upper ends 
of the curves in Fig.~\ref{FIG:Bifall}(b),(d). $M$ and $\omega$ are related to
each other almost linearly as in TWs \cite{LBBFHJ98}. 

The blow-up of the lower 
part of Fig.~\ref{FIG:Bifall}(b) in Fig.~\ref{FIG:Bifall}(c) shows how stability 
and shape of the solution branches change with $\psi$; to that end they are 
shifted such as to fit into Fig.~\ref{FIG:Bifall}(c).
While the SW at $\psi$=-0.03 has only one saddle-node the curvature of the 
branches changes with decreasing $\psi$ such that two additional saddle-nodes 
arise (for $\psi=-0.25,-0.3,-0.35$) with associated stability changes. 
For $\psi=-0.4$ we have only one saddle-node again. While the saddle-node 
positions slightly depend on the number of Galerkin modes retained the 
$r$-range with stable SWs [solid lines in Fig.~\ref{FIG:Bifall}(c)] definitely 
increases with decreasing $|\psi|$.

\begin{figure}
\centerline{\includegraphics[height=12.5cm,width=8cm]{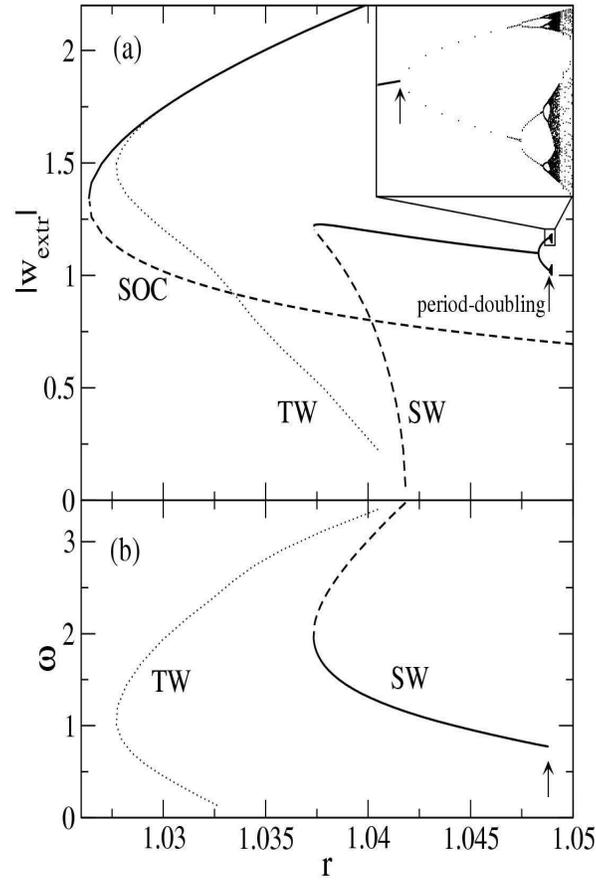}}
\caption{Details of the bifurcation behavior for $\psi=-0.03$
(leftmost curves in Fig.~\ref{FIG:Bifall}): (a) Magnitude
$|w_{extr}|$ of the extrema in the vertical flow. (b) Frequency $\omega$.
For the sake of completeness we include also the TW solution for laterally 
periodic boundary conditions allowing free phase propagation. 
SW and TW bifurcate subcritically at $r_{osc}=1.0418$ with Hopf frequency 
$\omega_H=3.426$. The TW branch ends by merging with zero 
frequency with the SOC solution branch. The SW solution becomes  
stable (solid lines) at the saddle-node position $r_s^{SW}=1.0373$.
At  $r=1.04831$ the MTS is broken and the solid SW line in (a)
splits into two when the magnitudes of the vertical flow extrema
occurring during one oscillation cycle become different [see, e.g., 
Fig.~\ref{FIG:wCMt}(b) where the downflow at $x=0=z$ is more intense than the
upflow]. This MTS-broken SW starts to undergo at $r = 1.04883$ 
(marked by arrows) a period-doubling route to chaos that is shown in more 
detail in the inset of (a).} 
\label{FIG:Bif003}
\end{figure}

The bifurcation behavior of the leftmost curves of Fig.~\ref{FIG:Bifall} is
displayed in more detail in Fig.~\ref{FIG:Bif003}. For the sake of completeness 
the TW solution (dotted curves) is shown as well. It bifurcates subcritically 
as an unstable solution out of the conductive state at the common SW-TW Hopf 
threshold. The TW solution ends by merging with zero frequency with the SOC 
branch. However, when phase propagation is suppressed as in our case then the 
TW does not exist and the upper SOC solution branch [full line in 
Fig.~\ref{FIG:Bif003}(a)] is stable down to its saddle-node. 

Also the SW becomes stable via a saddle-node bifurcation. With increasing 
heating 
$r$ the flow amplitude of the stable SW (full lines) slightly decreases. At 
$r=1.04831$ the MTS breaks and the solution branch splits into two. Thereafter
the downflow (upflow) extrema occurring in the SW oscillations, say, at $x$=0
($\pm \lambda/2)$ are more intense than the upflow (downflow) extrema.
Consequently, the time averaged fields have now a net SOC-like structure 
with non zero mean downflow (upflow), say, at $x$=0 ($\pm\lambda/2)$.

\begin{figure}
\centerline{\includegraphics[height=12.5cm,width=8cm]{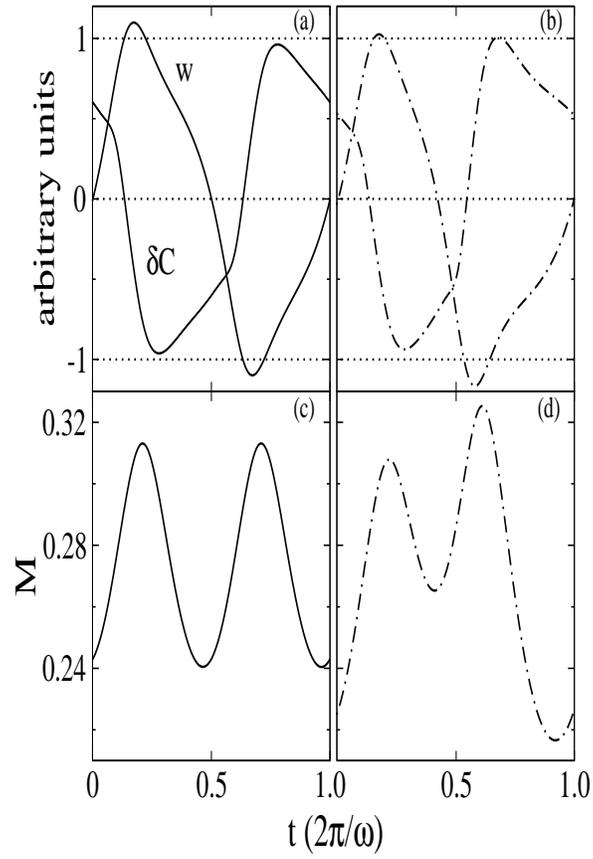}}
\caption{Effect of MTS-breaking on temporal oscillation profiles. Right (left) 
column shows a 
period-1 SW for $\psi=-0.03$ at $r=1.0488$ (1.0483) where the MTS is (not yet)
broken. Here $w$ and $\delta C$ are evaluated at midheight between two
rolls, $x=0=z$. The mixing number $M$ oscillates with twice the SW frequency 
as long as the MTS holds (c).}
\label{FIG:wCMt}
\end{figure}

Fig.~\ref{FIG:wCMt} shows the local dynamics of $w$ and $\delta C$ at $x=0=z$ 
and the global mixing number $M$ before (left column) and after (right column)
MTS breaking. By definition $M$ oscillates with twice the SW frequency as
long as MTS holds. Note that in particular $\delta C$ displays the
characteristics of a relaxational oscillator. In the MTS-broken SW of 
Fig.~\ref{FIG:wCMt} the extrema of upflow and downflow at $x$=0 differ. Also
the up- and downflow times between the respective zero crossings of $w$ differ.

\begin{figure}
\centerline{\includegraphics[height=12.5cm,width=8cm]{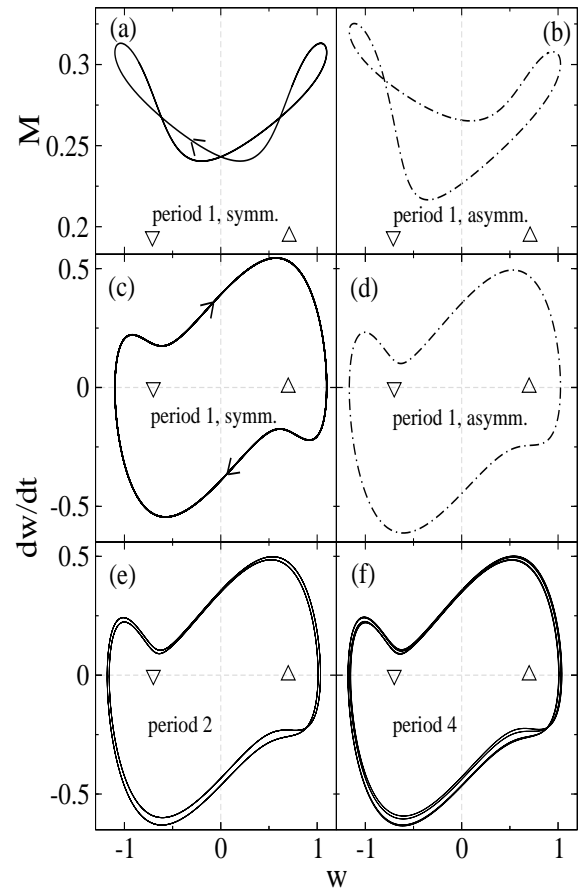}}
\caption{MTS-breaking and period doubling in the phase space dynamics of SWs.
Shown are the mixing number $M$ and $\dot w$ versus the vertical velocity $w$
at $x=0=z$. 
In (a)-(d) the dash-dotted (full) lines
refer to the period-1 SW in the right (left) column of
Fig.~\ref{FIG:wCMt} for which the MTS is (not yet) 
broken. Period doubling is displayed in (d) - (f). Upwards and downwards 
pointing triangles indicate symmetry degenerate unstable SOC fixed 
points (dashed SOC branches in Figs.~\ref{FIG:Bifall} and 
\ref{FIG:Bif003}) with upflow or downflow, respectively, at $x=0$.} 
\label{FIG:phaseplane}
\end{figure}

In Fig.~\ref{FIG:phaseplane} we show how MTS breaking and period-doubling
changes the SW phase dynamics using $w,\dot w$, and $M$ as characteristic
local and global quantities, respectively. The curves in 
Fig.~\ref{FIG:phaseplane}(a)-(c) refer to those of Fig.~\ref{FIG:wCMt}(a)-(c).
The upwards and downwards pointing triangles denote the two symmetry 
degenerate unstable SOC fixed points with either upflow or downflow (of
equal magnitude) located at $x=0$. The particular MTS-broken SW orbits of 
Fig.~\ref{FIG:phaseplane} move closer to the SOC with downflow at $x=0$. Here 
it would be interesting to see whether and how the heteroclinic orbits
connecting these two unstable symmetry degenerate SOC fixed points organize 
and restrain in phase space the dynamics of the SWs that periodically switch 
between up- and downflow.

For the $\psi$-range considered here we found that slightly after the MTS 
breaking a period-doubling scenario into chaos starts
that is compatible within our numerical resolution with the Feigenbaum
constant. For stronger Soret coupling, e.g. $\psi=-0.25$, we could resolve also
a $r$-window with period-3 SW states and subsequent period doubling. However,
we did not observe SWs beyond the chaotic window(s) seen, e.g., in the inset of
Fig.~\ref{FIG:Bif003}(a). After increasing the heating beyond this threshold 
the SWs developed transients into a stable SOC state 
with large convection amplitude [full line in Fig.~\ref{FIG:Bif003}(a)].

In summary, we have determined SW states in mixtures for several Soret coupling
strengths. Close to onset of convection the subcritical solutions are unstable.
Under phase-pinning conditions they become stable via saddle-node bifurcations.
After the occurrence of a MTS breaking they undergo a period-doubling cascade 
into chaos thereby terminating. 
 
\clearpage

 \clearpage



\begin{thebibliography}{999}
\bibitem{coupling}
For 5 weight \% of ethanol mixed into water at $T=20^{\circ} C$ the separation
ratio measuring the Soret coupling strength between concentration and
temperature gradients is $\psi \simeq -0.3$.
\bibitem{control}
We drive the system towards the unstable limit cycle (ULC) with frequency 
$\omega_\ast$ that is located on the subcritical SW solution branch as follows:
To reduce deviations with growing (decreasing) convection 
amplitude $A$ that decrease (increase) $\omega$ at fixed $R$ we set 
$A_{n+1}=A_n [1 + W_A(\omega - \omega_\ast)]$. Furthermore, when 
$\Delta \omega / \Delta t$ is negative (positive) we move towards  
the SW solution branch by decreasing (increasing) $R$ according to 
$R_{n+1}=R_n (1 + W_R \Delta \omega / \Delta t)$. The control depends on the
actual values of $\omega - \omega_\ast$ and $\Delta \omega / \Delta t$ and
vanishes only at the ULC with the preselected frequency $\omega_\ast$. The 
optimal weight factors $W_A$ and $W_R$ depend on $\omega_\ast$.
\end{thebibliography}
\end{document}